# Observation and investigation of narrow optical transitions of $^{167}$Er$^{3+}$ ions in femtosecond laser printed waveguides in $^7$LiYF$_4$ crystal

**Minnegaliev M.M.[1], Dyakonov I.V.[2], Gerasimov K.I.[1], Kalinkin A.A.[2], Kulik S.P.[2], Moiseev S.A.[1], Saygin M.Yu.[2], Urmancheev R.V.[1]**

[1] Kazan Quantum Center, Kazan National Research Technical University n.a. A.N. Tupolev-KAI, 10 K. Marx St., 420111, Kazan, Russia

[2] M.V. Lomonosov Moscow State University, Faculty of Physics, 1-2, Leninskie Gory, 119991, Moscow, Russia

**Abstract**. We produced optical waveguides in the $^{167}$Er$^{3+}$:$^7$LiYF$_4$ crystal with diameters ranging from 30 to 100 $\mu$m by using the depressed-cladding approach with femtosecond laser. These waveguides were studied (both inside and outside) by stationary and coherent spectroscopy on the 809 nm optical transitions between the hyperfine sublevels of $^4$I$_{15/2}$ and $^4$I$_{9/2}$ multiplets of $^{167}$Er$^{3+}$ ions. It was found that the spectra of $^{167}$Er$^{3+}$ were slightly broadened and shifted inside the waveguides compared to the bulk crystal spectra. We managed to observe a two-pulse photon echo on this transition and determined phase relaxation times for each of waveguides. The experimental results show that the created crystal waveguides doped by rare-earth ions can be used in optical quantum memory and integrated quantum schemes.

## 1. Introduction

Optical quantum memory (QM) is the key element for the creation of quantum repeater [1], which is a device that could provide a large distance trusted quantum key transfer in quantum communication systems. Such QM devices should be integrated in the existing telecommunication systems based on the optical waveguides. This requirement initiates the elaboration of the impedance matching waveguide QM schemes where the photon echo approach promises a convenient way for the storage of multi-qubit light fields [2, 3, 4]. Recent experiments have demonstrated highest quantum efficiency for gaseous and solid state media [5, 6, 7]. Herein, we note the proposals where QM cell is placed in high-q resonators [8, 9, 10, 11].

In the case of solid-state QMs, rare-earth ions (REI) in dielectric crystals are especially interesting for practical implementation. In particular, the optical transitions between multiplets of rare-earth ions within the 4$f$ shell have large coherence times and narrow inhomogeneous broadening [12, 13]. Moreover rareearth ions can be also used for long-lived storage on electron-nuclear spin transitions [14].

The off-resonant Raman protocol is promising for such storage and for the creation of a tunable addressed QM [15, 16, 17, 18]. One of the main requirements for achieving high efficiency in this protocol is small inhomogeneous broadening on the optical transition. The REI in the $^7$LiYF$_4$ crystal exhibits one of the smallest (for solid state systems) inhomogeneous broadenings of the absorption lines [19, 20, 21, 22, 23], which makes this crystal a good candidate for the offresonant QM schemes.

Recently a few new possible setups for integral waveguide QM schemes have been proposed as well. Authors of [24] demonstrated a storage of entangled photon pair and preservation of time qubits in 26 spectral modes [25] in a Ti:Tm:LiNbO₃ waveguide fabricated by $Ti^{4+}$ in-diffusion [26, 27, 28]. Considering the recent spectroscopic investigation of such waveguides at lower temperatures of the order of 0.8K [29], this approach shows good prospects for the creation of integral quantum repeaters. An alternative method for constructing of integral circuits has been proposed in [30]. This method is based on the fabrication of a planar waveguide structure on the surface of Pr:YSO crystal, where the active ions are accessed by evanescent field that extends into the crystal. A commercially available silicon optical fiber doped with erbium ions is also studied for using in integral QM [31, 32, 33]. Another approach towards the integral QM [34, 35] uses a photonic-crystal cavity fabricated in a Nd:YVO crystal and exploits focusedion-beam milling techniques [36]. Based on the system of multi-resonator, it is possible to implement the broadband QM integrated with a single optical waveguide [37]. Recently, an AFC protocol of photon echo quantum memory has been demonstrated in a waveguide written in Pr:YSO crystal [38] by using femtosecond laser writing approach. However the obtained experimental results for the waveguide QM schemes should be considerably improved in terms of the basic parameters of quantum storage before they can be used in any real application. That pushes further search for new crystals and waveguide QM systems.

Femtosecond laser writing (FSLW) is an established tool for rapid and cheap fabrication of 3D integrated photonic structures in an extremely wide range of optical materials [39] applied in many optical labs. In the original work [40], authors demonstrated waveguide writing capabilities in various optical glass substrates and further research [41, 42, 43, 44, 45, 46] extended the technology to crystals, ceramics and polymers. The refractive index change in crystals can be either negative or positive depending on the sample exposure conditions, thus some materials require cladding writing instead of the direct writing of the waveguide core. It should be noted that the waveguides were already harnessed in LiYF₄ crystal by this method [47, 48]. Such waveguides were then used to create an integrated circuit laser at room temperature [49]. Knowledge of the basic spectroscopic parameters of REIs in such waveguides, including homogeneous and inhomogeneous broadening of resonant lines, is necessary for the creation of integrated QMs and requires special careful study.

In this work, we created the waveguide structures in a $^{7}LiYF_4$:$^{167}Er^{3+}$ crystal by using the femtosecond laser writing method and we investigated the spectroscopic properties of these waveguides. Unique narrow optical transitions of $Er^{3+}$ ions give an opportunity for investigation of additional weak distortions introduced by crystal stressing during waveguide fabrication. The presented material is organized as follows. In Section 2, the manufacturing process and the geometric parameters of the waveguides in a crystal are described. Section 3 presents the results of stationary and coherent optical spectroscopy of waveguide structures with various diameters that were obtained at a temperature T = 4K. In Section 4 we discuss these results in the framework of the integral QM implementation.

## 2. Femtosecond laser waveguide writing in $^{7}LiYF_4$:$^{167}Er^{3+}$ crystal

The experimental setup of waveguides fabrication is shown in fig.1 [50]. The sample is exposed to a 400 fs pulse train emitted by a femtosecond fiber laser (Menlo Systems BlueCut) at 1 MHz repetition rate. The laser output is frequency doubled to meet optimal focusing conditions with a Mitutoyo Plan APO 100X microscope objective. Low-loss waveguide writing requires stable exposure conditions hence we implemented laser output power stabilization system driving the acousto-optical modulator at the output of the laser head based on the feedback signal generated by the photodiode. The 15.8x5x4 mm inorganic $^{7}LiYF_4$:$^{167}Er^{3+}$ crystal sample is mounted on the air-bearing positioner (Aerotech Fiberglide 3D). The sample's top surface is optically polished prior to waveguide writing process. The post-processing of the sample also requires optical polishing of the end-faces to ensure optimal light coupling conditions.

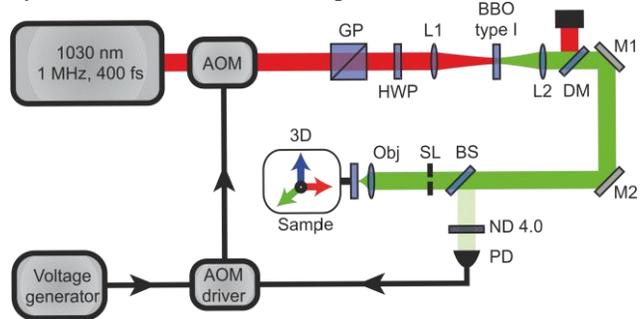

Figure 1. The femtosecond laser writing setup.

Spectroscopic applications require the waveguide core to possess properties close to the bulk crystal, thus writing the cladding and leaving the core with minimal damage to the crystal structure is beneficial for this purpose, hence we chose the type III waveguide writing geometry. Another advantage of this type of waveguides is the possibility of guiding a light with an arbitrary polarization. This design relies on mechanical stress induced refractive index increase in the core region governed by surrounding FSLW written damage tracks. Such waveguides were originally demonstrated in [51]. We have chosen circular geometry of the waveguide cladding, however depth dependent abberations distort the cladding to a slightly elliptical shape. Optimal cladding geometry could be engineered by maximizing overlap between waveguide eigenmode field pattern and the input field distribution. We have printed structures in $^7LiYF_4$:$^{167}Er^{3+}$ sample with 30, 50, 75 and 100 um transverse diameter (see fig.2). Each cladding track was written by translating the sample at 0.1 mm/s feed rate exposed by a laser beam with 94 nJ pulse energy at 1 MHz repetition rate with the polarization of the writing beam set parallel to waveguide axis. The cladding geometry parameters are summarized in Table1.

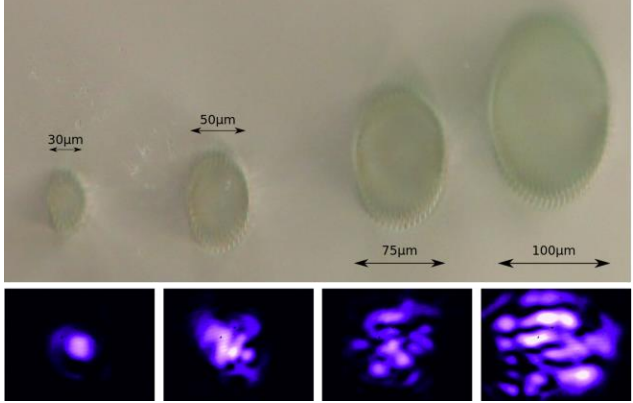

Figure 2. Waveguide end faces microscope image. Cross-section near-field image of beam propagated through the waveguides, pumping at 808nm.

Table 1. Cladding geometry parameters.

| N | Waveguide diameter, $\mu m$ | Inscription depth, $\mu m$ | Number of cladding tracks |
|---|---|---|---|
| 1 | 30 | 75 | 20 |
| 2 | 50 | 75 | 35 |
| 3 | 75 | 100 | 50 |
| 4 | 100 | 125 | 65 |

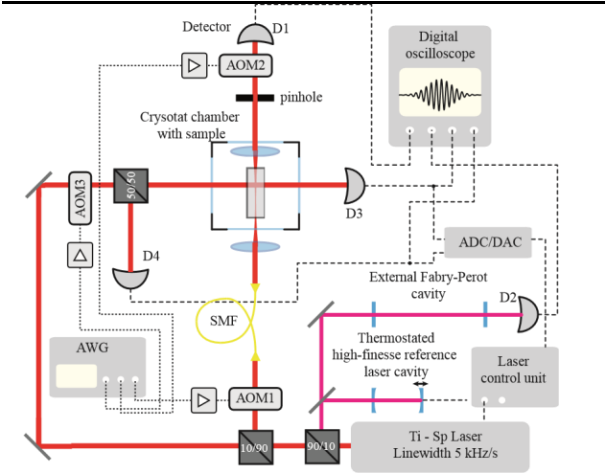

Figure 3. Experimental setup for the stationary and coherent spectroscopy of the waveguides in $^{167}Er^{3+}$:$^7LiYF_4$ crystal (Ddetector, AWG-arbitrary waveform generator, AOM-acoustooptic modulator, SMF-single mode fiber).

## 3. Optical spectroscopy of $^{167}Er^{3+}$ in the crystal waveguides

Fig.3 shows the simplified experimental setup for the stationary and coherent spectroscopy of the waveguides in $^{167}Er^{3+}$:$^7LiYF_4$ (0.005 at. %) crystal. Single frequency continuous Ti:Sp laser (Technoskan TIS-SF-777) was tuned to the $^4I_{15/2}(\Gamma_{56}) \rightarrow ^4I_{9/2}(\Gamma_{78})$ of

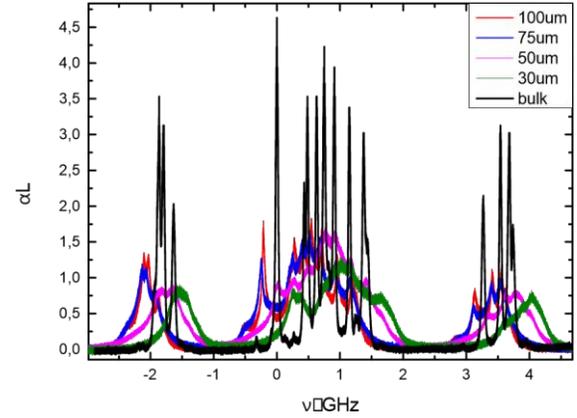

Figure 4. Absorption spectra of $^{167}Er$ in the bulk and waveguides with different diameters in $^{167}Er^{3+}$:$^7LiYF_4$ (0.005 at.%) crystal corresponding to the $^4I_{15/2}(\Gamma_{56}) \rightarrow ^4I_{9/2}(\Gamma_{78})$ transition. Zero on the frequency scale corresponds to the $\nu$=370.573 THz. T~4K. Electric field vector of the incident light field was parallel to the crystal's $C_4$ axis.

$^{167}Er^{3+}$ ion transition corresponding to the 809 nm wavelength. First, we performed the stationary spectroscopy. The light was launched into the crystal from two orthogonal sides through AOM1 and AOM3 which were both open and we scanned the frequency by laser built-in resonator. An external Fabry-Perot cavity with a free spectral range of 354.4 MHz was used to control and to correct the nonlinear frequency scanning of the laser.

We used single mode fiber for spatial filtering of the light beam after AOM1. Then the light mode was focused by the lens with the focal length $f = 25.4 mm$ located outside of the cryostation chamber. The crystal was placed inside the closed-cycled cryostat (Montana Instruments) at ~4 K temperature on a high precision translation stages (Attocube ANPx101/LT) which allowed us to move the crystal in x-y plane (parallel to optical table). Another lens was located inside the cryostat on the distance of its focal length ($f = 35 mm$) from the end face of the crystal for collecting the emission from the waveguide. After the cryostat, the pinhole was introduced to select only the light beam passed through the waveguide. That way detector D1 (Thorlabs PDA120A/M) gathered the light spectrum from the specific waveguide while detector D3 simultaneously saved the bulk spectrum to see the frequency shift between the two spectra and AOM2 was always open in this experiment. Here and in what follows, the AOMs were controlled using an arbitrary wave-form generator (Rigol

DG5352). We acquired the signals with a digital oscilloscope (Tektronix DPO 7104C).

The observed absorption spectra for bulk crystal and waveguides printed in the $^7LiYF_4$:$^{167}Er^{3+}$ crystal are illustrated in fig.4. Frequency origin corresponds to 370.573 THz transition. Light polarization was parallel to crystal's $C_4$ axis. The absorption lines of ions inside the waveguide are broadened due to additional spatial inhomogeneities introduced by the local tensions caused by the waveguide creation process. For example the inhomogeneous width of the narrowest and strongest absorption line for bulk crystal was 24 MHz. The optical density falls subsequently since the atomic absorption lines are spread over the wider spectral range. The interesting fact is the shifting of the whole spectrum in comparison with the spectrum in the bulk crystal. So spectrum observed for 100 and 75 $\mu$m waveguides shifts to the lower frequencies, but the spectrum for 30 $\mu$m waveguide - to the higher frequencies. However, there is no significant shift for spectrum of 50 $\mu$m waveguide in comparison to the spectrum in bulk crystal.

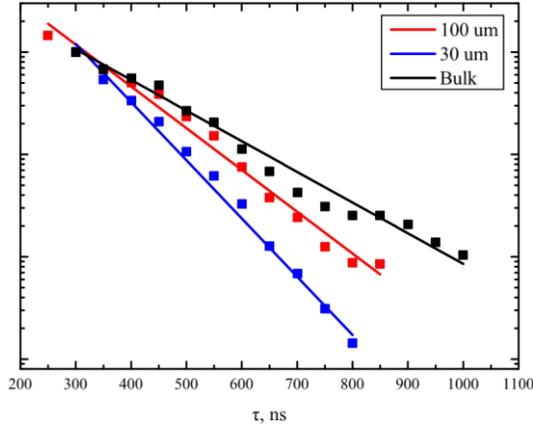

Figure 5. Two-pulse photon echo signal decay as a function of time delay between the light pulses for bulk (black) and for 100 and 30 $\mu$m waveguides (red and blue, respectively) waveguides in $^{167}Er^{3+}$:$^7LiYF_4$ (0.005 at.%) crystal on the $^4I_{15/2}(\Gamma_{56}) \rightarrow {}^4I_{9/2}(\Gamma_{78})$ transition. T ~ 4K.

For the second experiment we studied a two-pulse (primary) photon echo in the fabricated waveguides that we previously observed in bulk crystal [23] . This time AOM1 carved the pulses from the continuous laser emission and AOM2 was used for protection of the sensitive avalanche detector D1 (Thorlabs PDA120A/M) from intense input pulses and allowed light to pass only during the echo signal measurement. In this case, the optical channel with AOM3 was used for long-term laser frequency stabilization as in work [23]. All photon echo signals were measured on the most intense line of absorption spectra located near the zero on the frequency scale of fig.4. The phase relaxation time $T_2$ of the studied optical transition was measured in primary photon echo experiments. Herein, we used 100-ns duration Gaussian pulses. We determined the pulse duration as full width at half maximum. The measurement repetition rate was determined by the operation cycle frequency of the cryostat and was ~ 0.8 Hz.

Table 2. Phase relaxation $T_2$ times for bulk and waveguides with different diameters in $^{167}Er^{3+}$:$^7LiYF_4$ (c=0.005 at.%) crystal of the $^4I_{15/2}(\Gamma_{56}) \rightarrow {}^4I_{9/2}(\Gamma_{78})$ transition. at T ~ 4K.

| Waveguide diameter, $\mu$m | $T_2$, ns | Error, ns |
| --- | --- | --- |
| bulk | 580 | 17 |
| 100 | 425 | 13 |
| 75 | 388 | 19 |
| 50 | 377 | 11 |
| 30 | 305 | 9 |

Despite to the broader absorption lines compared to the bulk crystal and consequently smaller echo efficiencies, we observed the primary echo in all the waveguides where $T_2$ times were measured for each of them. Two pulse photon echo signal decay versus time delay between the laser pulses for the bulk crystal and for 100 and 30 $\mu$m crystal waveguides is depicted in fig.5. Each data point in the graphs of fig.5 was obtained by averaging the echo signal intensity over 100 measurements. Obtained $T_2$ times for bulk crystal and for each waveguide are presented in the Table 2. The difference in $T_2$ times obtained for the bulk crystal in this work and in recent work [23] is caused by slightly higher temperature in this experiment. One can observe from Table 2 that the phase relaxation time decreases from $T_2 = 580 \pm 17$ ns for bulk crystal down to $T_2 = 305 \pm 9$ ns for 30 $\mu$m diameter waveguide.

## 4. Conclusion

In conclusion, we fabricated optical waveguides with diameters ranging from 30 $\mu$m to 100 $\mu$m in $^{167}Er^{3+}$:$^7LiYF_4$ crystal by femtosecond laser writing method. These waveguides can guide the light with arbitrary input polarization. The performed stationary optical spectroscopy have shown relatively small additional broadening (less than 200 MHz) of the absorption lines with a decrease of the waveguide diameter. Also we observed two-pulse photon echo in all the fabricated crystal waveguides. The optical phase relaxation time indicated a small sensitivity to the waveguide diameter. Our study shows good perspectives for implementiation of the integrated optical QMs in the investigated crystal waveguides due to the small changes of the basic spectroscopic parameters so important for many QM protocols.

The reported study was funded by RFBR according to the research projects no. 17-02-00918 and no. 17-52-560009.